\def\bq{\begin{equation}}
\def\eq{\end{equation}}
\def\bqa{\begin{eqnarray}}
\def\eqa{\end{eqnarray}}
\def\bqb{\begin{eqnarray*}}
\def\eqb{\end{eqnarray*}}
\documentstyle[12pt]{article}\textwidth 16cm
\def\pr#1#2#3{ Phys. Rev. ${\bf{#1}}$ (#2) #3}

\def\pl#1#2#3{ Phys. Lett. ${\bf{#1}}$ (#2) #3 }

\def\np#1#2#3{ Nucl. Phys. ${\bf{#1}}$ (#2) #3}

\newcommand{\vv}{\vspace{4.mm}}
\newcommand{\Lam}{\Lambda }
\newcommand{\epsi}{\epsilon }
\newcommand{\ZP}{$Z'\ $}

\global\nulldelimiterspace = 0pt
 
\def\Bsl{\hbox{/\kern-.6700em$B$}} 
\def\Dsl{\hbox{/\kern-.6700em$D$}} 
\def\Wsl{\hbox{/\kern-.6700em$W$}} 

\def\roughly#1{\mathrel{\raise.3ex
    \hbox{$#1$\kern-.75em\lower1ex\hbox{$\sim$}}}}

\def\mh2{m^2_H}

\begin{document}
\pagenumbering{arabic}
\thispagestyle{empty}
\hspace {-0.8cm} CPT-96/P.3371\\
\hspace {-0.8cm} July 1996\\
\vspace {0.8cm}\\
 
\begin{center}
{\Large\bf  HERA PROSPECTS ON COMPOSITENESS AND NEW VECTOR BOSONS} \\

 \vspace{1.8cm}
{\large  P. Chiappetta$^a$ and J.-M. Virey$^{a,b,1}$}
\vspace {1cm}  \\
$^a$\,Centre
de Physique Th\'{e}orique,
UPR 7061,\\
CNRS Luminy, Case 907, F-13288 Marseille Cedex 9.\\
\vspace{0.5cm}

$^b$\,and Universit\'e de Provence, Marseille, France.\\  
\vspace{2.5cm}

 {\bf Abstract}
\end{center}
\noindent
The absence of deviations from the Standard Model for the
differential cross section $\frac{d\sigma}{dQ^2}$ at HERA is used to
set limits on electron quark compositeness scale and on new vector
bosons, especially the hadrophilic one recently introduced as a
possible explanation for LEP/SLC and CDF
anomalies.

\vspace{1cm}
\vfill
\begin{flushleft}

------------------------------------\\
{$^1$} Moniteur CIES and allocataire MENESR. \\
email : chiapeta@cpt.univ-mrs.fr

\end{flushleft}

\setcounter{page}{0}
\def\thefootnote{\arabic{footnote}}
\setcounter{footnote}{0}
\clearpage
\section{Introduction.} 

The $e^{\pm}p$ HERA collider, in which electrons and
positrons of $27.5$ GeV of energy collide with $820$ GeV protons, can
extend the search for compositeness \cite{compo} in the
electron-quark channel or new gauge bosons \cite{CGZ} far beyond the
kinematical limit thanks to indirect effects. The observable we will
deal with is the differential cross section   $\frac{d\sigma}{dQ^2}$ ,
where $Q^2$ is the positive squared transfer momentum. 
The purpose of this letter is to set limits on electron quark 
compositeness and new vector gauge bosons from present HERA 
measurements\cite{H1} and future prospects.  
\section{Bounds from present data.} 

The first
analysis, we will perform,  relies on purely inclusive  $H1$ 
measurements\cite{H1} in deep
inelastic scattering from $ep$ collisions with an integrated luminosity
of $0.909 pb^{-1}$ for electron and $2.947 pb^{-1}$ for positron beams.
The measured $Q^2$ range lies between $200 GeV^2$ and $2. 10^4 GeV^2$. A
$\chi^2$ analysis of $e^{\pm}p$ differential cross sections - including
statistical and systematical errors added in quadrature  except for an
overall normalization uncertainty- is performed. The
overall normalization errors are $3.5\%$ (resp $1.8 \%$) for $e^-p$
(resp $e^+p$) data. 
%
%
%

If quarks and leptons have a substructure, the contact
interaction, on which our derivation is based,  is given by the 
lagrangian \cite{compo}:

\bqa  L_{NC} = \sum_q ( \eta_{LL}(\bar e_L \gamma_{\mu} e_L) 
(\bar q_L \gamma^{\mu} q_L) +  \eta_{RR}(\bar e_R \gamma_{\mu} e_R) 
(\bar q_R \gamma^{\mu} q_R)  \nonumber 
  \\   + \eta_{LR}(\bar e_L
\gamma_{\mu} e_L)  (\bar q_R \gamma^{\mu} q_R) + \eta_{RL}(\bar e_R
\gamma_{\mu} e_R)  (\bar q_L \gamma^{\mu} q_L) )  \eqa
 
where $\eta_{IJ}= \epsilon \frac{g^2}{\Lambda^2}$ (with
$\epsilon=\pm1$, from now $\Lam$ means $\Lam_{eq}$). 
The present HERA limits, which can be inferred from the
absence of deviations from SM, are given in the following table
(assuming $g^2=4 \pi$). 

\vv
\begin{center}
\begin{tabular}{|c||c|c|c|c|}
\hline
$\Lam$ (TeV)&$\Lam_{LL}$&$\Lam_{RR}$&$\Lam_{LR}$&$\Lam_{RL}$\\
\hline
\hline
$\epsi = +1 $ & $ 2.3  $ & $ 2.3  $ & $ 2.5  $ & $ 2.5  $ \\
\hline
$\epsi = -1 $ & $ 1.0  $ & $ 1.0  $ & $ 1.2  $ & $ 1.2  $ \\
\hline
\end{tabular} 
\end{center}
\begin{center}
Table 1: Present HERA limits on $\Lam$ at 95\% CL \cite{H1}.
\end{center}

  In fact, the world present limits on $\Lam$ come from the search for
an excess of high mass dileptons events at Tevatron \cite{CDFlim},
which gives for a
data sample of $L= 110 pb^{-1}$ (Runs Ia \& Ib) the values $\Lam^+\geq$ 2.6 TeV
and $\Lam^-\geq$ 3.8 TeV, where the sign + or $-$ corresponds to $\epsi$.

\vspace{8.mm}

 At low energies, compared to the mass of the heavy
particle, this lagrangian can also describe the effect of a vector boson
exchange. 

The $Z'$ current, written in terms of left handed $C'_L$ and right 
handed $C'_R$ couplings to fermions,  reads:

\bq J_{\mu}^{Z'}= \frac{1}{\sin \theta_W \cos \theta_W} \bar f
\gamma_{\mu} ( C'_L \frac{(1-\gamma_5)}{2} + C'_R \frac{(1+\gamma_5)}{2}
) f \eq

 In order to fix the normalization we also give the fermionic
couplings to the standard model Z : $C_L=I_3-Q \sin^2 \theta_W$
(with $I_3= \pm \frac{1}{2}$) and  $C_R=-Q \sin^2 \theta_W$, Q being
the fermion electric charge and $\theta_W$ the Weinberg angle (
$\sin^2 \theta_W = 0.2319$).

\vv

The previous compositeness limits given in table 1 can  be used to 
set limits on the $Z'$ couplings to fermions for a fixed $Z'$ mass, 
according to:

\bq C'_e C'_q  \sim (\frac{\sin^2 \theta_W \cos^2 \theta_W}{\alpha})
(\frac{M^2_{Z'}}{\Lambda^2}) \eq
 Assuming typical values like $M_{Z'}= 1$ TeV and $\Lambda= 2.5$ TeV
 (i.e. the most favorable case) one gets:
\bq C'_e C'_q  \sim 4. \eq

\vv

Recently, the CDF collaboration\cite{CDF} has reported an excess of jets
at large transverse energy (more precisely for $E_{T} \geq
250$ GeV). A possible non standard -i.e. beyond SM- explanation could be
either existence of a quark substructure (at a scale $\Lam_{qq} \sim 1.6$
TeV), or existence of an extra heavy neutral vector boson, of mass
around 1 TeV. It has to  couple  dominantly to quarks, in order to
explain also the LEP/SLC anomalies\cite{RB} in the heavy quark sector
i.e. deviations from the SM on $R_b$ and $R_c$ \cite{PC} \cite{ALT}.
This $Z'$ provides an additional contribution to top
production\cite{STIR}, still compatible with present measurements. The
potentialities of LEP2 \cite{PC} and polarized RHIC\cite{JM} for the
observability of hadrophilic $Z'$ effects have also been explored. It
is of some importance to investigate if the preferred range for
hadrophilic $Z'$ couplings to fermions is ruled out by present data
from HERA. Without any
assumption on the underlying theory\cite{PC}  these couplings can be
expressed in terms of ratios $\xi_{Vf}$ and $\xi_{Af}$ between the $Z'$
couplings and the Z couplings. In the leptophobic option\cite{ALT} 
equality of the
left-handed couplings within one $SU(2)_L$ doublet, labelled by the
parameter $C'_L=x$,is imposed whereas the right-handed couplings 
$C'_{Rq}$ for up and
down type quarks (labelled as $y_u$ and $y_d$ in \cite{ALT}) are left free. The 
relationship between hadrophilic and leptophobic options is given by:

\bq \xi_{Vf}= \frac{C'_{Lf}+C'_{Rf}}{I_3-2Q_f\sin^2 \theta_W}\;\;,\;\;        
\xi_{Af}=\frac{C'_{Lf}-C'_{Rf}}{I_3} \eq 

The final fit is obtained for the parameters:
$x=-1, y_u=2.2, y_d=0$ \cite{ALT}. There is no direct $Z'$ coupling to leptons: 
it occurs only through the mixing between Z and $Z'$, which is small 
 ($\xi \sim 3.10^{-3}$). If the $Z'$ 
strictly does not couple to leptons, no restriction can be inferred 
from $e^+e^-$ and $ep$ colliders.
 For the hadrophilic case \cite{PC}, under the assumption that
the $Z'$ leptonic couplings are the same as the Z ones, only very large
$\xi_q$ values are excluded from eq(4) i.e. of the order of $30-100$. 
Since under the assumption that the $Z'$
cannot be wide  one gets: $\xi_u \sim \xi_d \leq 3-4$\cite{PC}, 
then the present HERA data are not sensitive to the hadrophilic \ZP
in the parameter range needed to accomodate LEP/SLC and CDF anomalies.

 \section{Restrictions from future data}

We shall now discuss the restrictions to be expected at HERA at $\sqrt
s=314$ GeV with two luminosity options, namely $L_1=1000 pb^{-1}$ and 
$L_2=500 pb^{-1}$ for electrons and positrons. Assuming a $Q^2$
resolution $\frac{\Delta Q^2}{Q^2} = 0.5$, table 2 gives the $95 \%$ CL
limits concerning the compositeness scale $\Lambda$, using a $\chi^2$ 
analysis. Statistical and systematical errors have been added in 
quadrature. Since the systematical error is small -roughly $2\%$ - we 
are dominated by the statistics in large $Q^2$ range we are interested 
into. 

The main sensitivity 
is obtained for up quarks. We have used the GRV parametrization of 
structure functions\cite{GRV} and checked that choice of other sets 
like \cite{BS}, \cite{MRS} leads to tiny deviations ($\sim 3 \%$). 
Our results are weakly sensitive to the $Q^2$ resolution.
\vv
\begin{center}
\begin{tabular}{c|c||c|c|c|c|}
\cline{2-6}
&$\Lam$ ( TeV)&$\Lam_{LL}$&$\Lam_{RR}$&$\Lam_{LR}$&$\Lam_{RL}$\\
\hline
\hline
$L_1=1.0 fb^{-1}$ & $\epsi = +1 $ & $ 8.55 $ & $ 8.4  $ & $ 8.0  $ & $ 7.9  $ \\
\cline{2-6}
& $\epsi = -1 $ & $ 8.4  $ & $ 8.2  $ & $ 7.7  $ & $ 7.6  $ \\
\hline
$L_2=0.5 fb^{-1}$ & $\epsi = +1 $ & $ 7.2  $ & $ 7.05 $ & $ 6.75 $ & $ 6.7  $ \\
\cline{2-6}
& $\epsi = -1 $ & $ 7.0  $ & $ 6.85 $ & $ 6.4  $ & $ 6.35 $ \\
\hline
\end{tabular} 
\end{center}
\begin{center}
Table 2: HERA limits on $\Lam$ at 95\% CL.
\end{center}

The best limits on the left left or right right chiralities are obtained 
from electron channel whereas the positron channel is more sensitive to 
left right and right left chiralities. Moreover the cross section from 
positron beam is characterized by a destructive interference 
between the photon and the massive gauge boson Z. These comments 
explain why the bounds on $\Lambda_{LR}$ and $\Lambda_{RL}$ in table 2 are 
slightly weaker than those on $\Lambda_{LL}$ and $\Lambda_{RR}$. The fact 
that present HERA data restrict more 
$\Lambda_{LR}$ and $\Lambda_{RL}$ is only due to the difference in 
luminosity between the electron and positron beams. 

\vv

  These values of $\Lam$ reachable at HERA have to be compared
to limits coming from others future or upgraded experiments. Firstly, Tevatron
with the run II corresponding to an integrated luminosity of $L= 2 fb^{-1}$,
can reach $\Lam \sim$ 5 TeV \cite{CDFlim}, and secondly, experiments measuring
parity violation in muonic atoms and cesium can give bounds on $\Lam$ above
10 TeV if parity is strongly violated \cite{atom}.

\vspace{8.mm}

The mass limits 
on several types of "conventional" new vector bosons (of $E_6$, left-right origin for
example)  are given in table 3. SSM refers 
to the sequential standard model, whereas $LR_S$ labels the 
symmetric left right model and ALR the alternative one
 ( for more details we refer to \cite{CGZ}).

\vv
\begin{center}
\begin{tabular}{|c||c|c|c|c|c|c|c|}
\hline
$M_{Z^{'}}$ ( GeV)& SSM &$ \chi $&$ \psi $&$ \eta $&$ \eta_{\bot} $&$ LR_s $
&$ ALR $\\
\hline
\hline
$L_1=1.0 fb^{-1}$ & $ 750  $ & $ 470  $ & $ 260  $ & $ 290  $ & $ 360  $
& $ 500 $ & $ 730  $\\
\hline
$L_2=0.5 fb^{-1}$ & $ 630  $ & $ 390  $ & $ 210  $ & $ 240  $ & $ 300  $
 & $ 420  $ & $ 610  $\\
\hline
\end{tabular} 
\end{center}
\begin{center}
Table 3: HERA limits on $M_{Z^{'}}$ at 95\% CL.
\end{center}

These bounds are not competitive with Tevatron ones since the present 
mass limits are around 600 GeV and will be of order 800 GeV for the run II
\cite{CDFlim}. We can note that low energy experiments, measuring atomic
parity violation, are also deeply sensitive
to the presence of new gauge bosons and can give limits $\geq$ 500 GeV, see
\cite{atom} for further details.

\vspace{8.mm}

Finally, we have constrained the parameter space of the hadrophilic \ZP of
ref. \cite{PC} under the asumption that the new couplings to leptons
are identical to the Standard one's.

\vv

We give in fig. 1 the restrictions for the hadrophilic $Z'$\cite{PC}
on the parameters $\xi_{Au}$ and $\xi_{Vu}$ with $M_{Z^{'}}=1$ TeV,  
setting $\xi_{Ad}=\xi_{Vd}=0$.
Now, if we allow the d-coupling to be non-zero there is, on one hand 
a decrease of the area of the ellipsis, and on the other hand a weak 
shift of the ellipsis along a direction which depends on the sign and 
the magnitude of the $\xi_{d}$ parameters. More precisely, if $\xi_{Ad}$
 increases towards positive (resp. negative) values the ellipsis moves
 in the direction of negative (resp. positive) $\xi_{Vu}$. The same 
behaviour holds for $\xi_{Vd}$ but with a weaker shift along the $\xi_{Au}$ axis.
For other values of $M_{Z^{'}}$ the scaling formula
of \cite{PC}i.e. $\xi_{Vq},\xi_{Aq} \sim \frac{M_{Z'}}{1 TeV}$ works well.\\
If the leptonic couplings are much weaker than the SM value, which 
was our working assumption, the allowed domain for $Z'$ couplings to up quarks 
is much larger. 

\vv
For the model of Altarelli and collaborators \cite{ALT},
the restriction obtained from eq(3), assuming 
$C'_e \sim \xi \sim 3. 10^{-3}$ leads to $C'_q \sim 200$. Therefore 
the leptophobic $Z'$ will escape HERA constraints. 

 \section{Conclusions}
Already with the unpolarized option the realistic future prospects 
for HERA will push compositeness scale in the electron quark sector in 
the range of $10$ TeV. This limit is twice larger than Tevatron
expectations and comparable to atomic parity violation prospects.
Mass limits on a large class of new vector bosons 
like those of $E_6$ and left right origin, are not competitive with 
Tevatron. On the other hand HERA can put stringent limits on the 
hadrophilic  $Z'$ couplings to quarks, recently advocated to explain 
LEP/SLC and CDF anomalies.  
 If in $E_6$ theories with orbifold compactification 
\cite{ORBI} or flipped $SU(5)$\cite{NANO} the leptophobic symmetry is 
realized, these $Z'$ will escape 
HERA tests. 

 \vspace{0.5cm}

{\bf Acknowledgements} \par
This work has been partially supported by the EC contract
CHRX-CT94-0579. We would like to thank Claudio Verzegnassi and F.M. Renard for discussions at an early stage of this work. We are indebted to Raoul Gatto for enlightening 
precisions on the leptophobic $Z'$ and to Marie Claude Cousinou, Claude 
Vall\'ee, Elemer Nagy, Chafik Benchouk and Laurent Lellouch 
for valuable advice on the $\chi^2$ analysis.

\centerline {\ {\bf Figure Caption }}

Fig.1 Allowed parameter space for model of ref. \cite{PC},
$\xi_{Au}$ vs $\xi_{Vu}$ with $M_{Z^{'}}=1. TeV$ and 
$\xi_{Ad}=\xi_{Vd}=0$. The plain ellipsis gives the allowed domain 
for $L_1=1.0 fb^{-1}$ (internal) and $L_2=0.5 fb^{-1}$ (external); 
the dashed ellipsis  gives the allowed domain corresponding to
$\Gamma_{Z^{'}} \leq 500 Gev$.\\


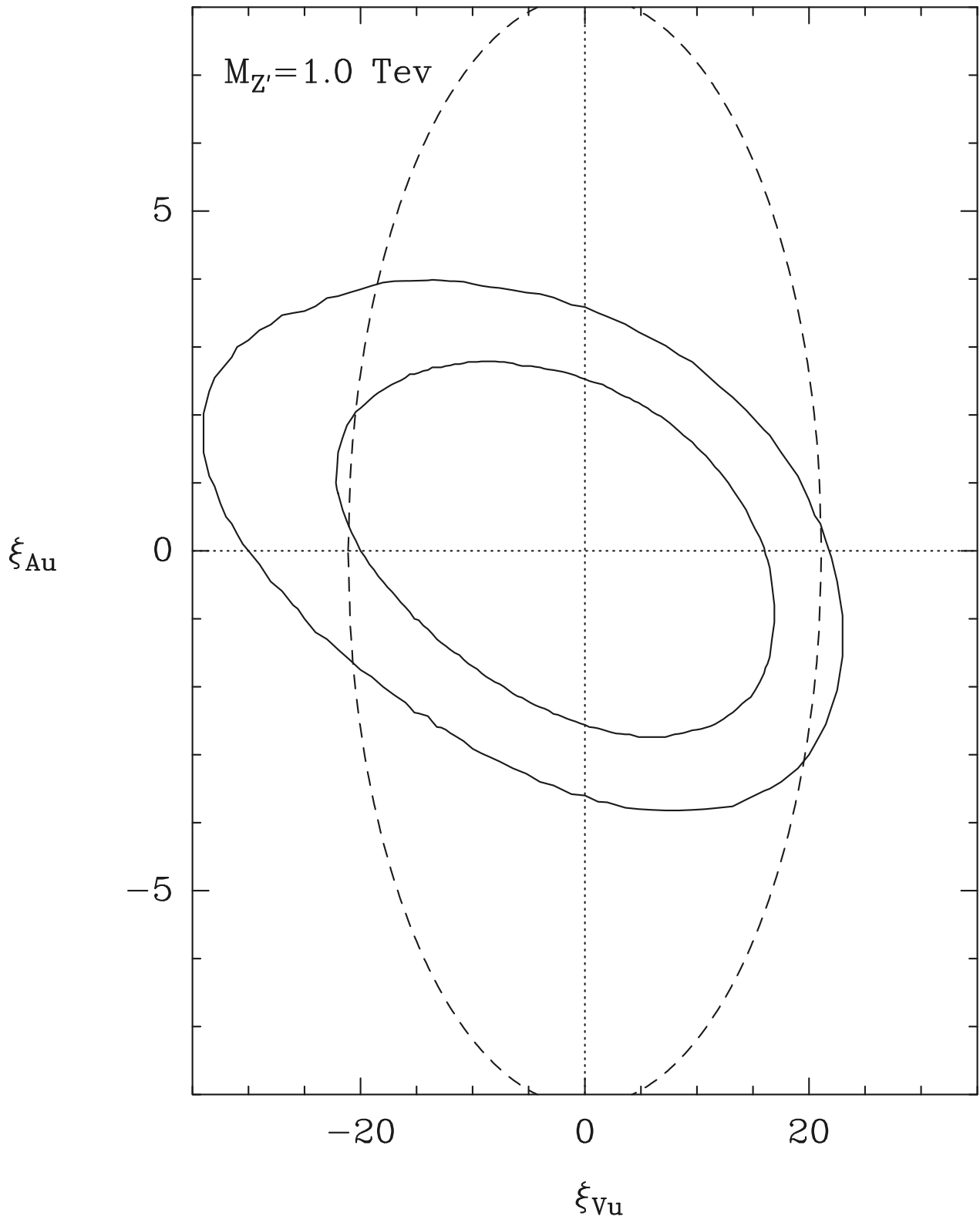
\end{document}